\documentclass[aps,prd,twocolumn,showpacs,groupedaddress,nofootinbib,amsmath]{revtex4}
\usepackage{graphicx}
\usepackage{epsfig}
\usepackage{multirow}
\usepackage{subfigure}
\newcommand{\be}{\begin{equation}}
\newcommand{\ee}{\end{equation}}
\newcommand{\ba}{\begin{eqnarray}}
\newcommand{\ea}{\end{eqnarray}}

\newcommand{\Dslash}{\lower-0.18ex\hbox{\makebox[-1pt][l]{\,/}}D}

\begin{document}

\title{Nature of the chiral phase transition of two flavor  QCD  from imaginary chemical potential with HISQ fermions  }

\author{Liang-Kai Wu}
\thanks{Corresponding author. Email address: wuliangkai@163.com}
\affiliation{Faculty of Science, Jiangsu University, Zhenjiang, 212013, People's Republic of China}
\affiliation{National Supercomputer Center in Wuxi, Wuxi, Jiangsu, 214072, People's Republic of China}
\affiliation{Department of Physics, College of Science, Swansea University, Swansea, SA2 8PP, United Kingdom}
\affiliation{Key Laboratory of Quark and Lepton Physics (MOE), Central China Normal University, Wuhan, 430079, People's Republic of China}

\author{Fa-Ling Zhang}
\affiliation{Faculty of Science, Jiangsu University, Zhenjiang, 212013, People's Republic of China}

\date{\today}

\begin{abstract}
The nature of the thermal phase transition of two flavor QCD in the chiral limit has an important implication for the QCD phase diagram.
We carry out lattice QCD simulations in an attempt to address this problem.
Simulations are conducted with
a Symanzik-improved gauge action and  the HISQ  fermion action.  Within the imaginary chemical potential formulation, five different quark masses,  $am=0.020,\, 0.018, \, 0.015, \,
0.013,\, 0.010$, and four different
lattice volumes  $N_s=8, \, 12,\,  16, \,  20$ with temporal extent $N_t=4$  are used to explore the scaling behavior. At each of the  quark masses, the Binder cumulants of the chiral condensate on different lattice
volumes  approximately intersect at one point.   We find that at the intersection point, the Binder cumulant $B_4(am,a\mu_c) $ is around $3$
which deviates from the $Z(2)$ universality class value 1.604.
However, based on the expectations of  $Z(2)$ criticality,   the fitting result only with the data from
the largest lattice volume $N_s=20$  agrees well with earlier result [ Phys. Rev., D90, 074030(2014) ]\cite{Bonati:2014kpa}. This fact
implies that, although the finite cut-off effects
could be reduced with HISQ fermions even on $N_t=4$ lattices,  larger lattices with spatial extent  $N_s>=20$ for such studies are needed
 to control finite volume effects.
\end{abstract}


\maketitle

\section{INTRODUCTION}
\label{SectionIntro}

The thermodynamics of matter described by QCD  is characterized by a transition from  the low-temperature hadronic phase with confined quarks and gluons to the high-temperature
phase with deconfined quarks and gluons. This phase transition  is relevant to the early Universe,
compact stars,  and heavy ion collision experiments.  Reviews on the study of the phase diagram can be found in Refs.~\cite{Fukushima:2010bq,Fukushima:2011jc,Aarts:2015tyj}
 and references therein. Mapping out the  phase diagram of QCD is one of the most challenging tasks presented for theoretical  physics.    Although substantial progress has been achieved
 in determining the phase diagram of QCD at zero density, the nature of the  phase transition of  QCD  with two massless flavors remains still open.

 At the transition point, if the $U_A(1)$ symmetry is not restored,  QCD  with two massless flavors has the symmetry-breaking pattern
 $[SU(2)_L \times SU(2)_R]/Z(2)_V \hspace{-1mm}\rightarrow SU(2)_V/Z(2)_V $,  on the other hand, if the $U_A(1)$ symmetry is effectively and fully  restored,  QCD  with two massless flavors has the symmetry-breaking pattern
 $[U(2)_L \times U(2)_R]/U(1)_V \rightarrow U(2)_V/U(1)_V $~\cite{Pisarski:1983ms,Butti:2003nu,Pelissetto:2013hqa}. For two-flavor QCD, Pisarsky and Wilczek pointed out that, if the $U_A(1)$ symmetry is broken
 at transition point $T_c$,  the system undergoes second-order transition with $O(4)$ scaling, although not necessarily. On the other hand, if the $U_A(1)$ symmetry is restored at
 $T_c$, the system undergoes a first-order transition. However, further studies~\cite{Butti:2003nu,Pelissetto:2013hqa} show that,  even if the $U_A(1)$ symmetry is restored at
 $T_c$, the system also may have an infrared stable fixed point, so the transition can be of either first order or second order with  different critical exponents from the O(4) universality class.
 Reference~\cite{Sato:2014axa} suggests the transition is of second order, but
 one of  critical exponents is different from the standard $O(4)$ model.

  As the interaction between the quarks and gluons is inherently
 strong at hadronic energy scales, lattice QCD simulation is the most reliable method up to date. The standard method to address the nature of QCD  with two massless flavors is to carry out
 simulations by successively reducing the quark mass, and in the meantime, monitoring the transition behaviour. If the transition is of second order in the chiral limit,
 then this second-order transition disappears immediately when finite quark masses are turned on. On the other hand, if the transition is of first order in the chiral limit, it will get weakened
 gradually until at a certain $Z(2)$ point as the quark masses increase.

 Considerable work using lattice QCD simulations has been devoted to this problem. Some lattice QCD simulation studies favor a second-order
 transition~\cite{Ejiri:2009ac,Aoki:1998wg,Karsch:1994hm,Iwasaki:1996ya,AliKhan:2000wou,Ejiri:2015vip,Bernard:1996iz,Burger:2011zc},
 some support that  the transition is of the first order~\cite{DElia:2005nmv,Bonati:2014kpa,Philipsen:2016hkv,Cossu:2013uua,Aoki:2012yj,Cuteri:2017gci}
 , and some  favor neither~\cite{Fukugita:1990dv,Bernard:1999fv}.
 For a discussion, see Refs.
 \cite{Meyer:2015wax,Bonati:2014kpa,Burger:2011zc} and references therein.

 Apart from the conventional method which focuses  on the critical exponents, the nature of
 the phase transition of QCD  with two massless flavors can be addressed  by exploring the fate of $U_A(1)$ symmetry at high
 temperature~\cite{Aoki:2012yj,Cossu:2013uua, Ohno:2012br,
Dick:2015twa,Brandt:2016daq,Tomiya:2016jwr,Bazavov:2012qja}. Reference~\cite{Cuteri:2017gci} discusses this problem from the aspect of
   noninteger number of flavors.
 In Ref.~\cite{Bonati:2014kpa}, a novel approach has been developed to address the nature the phase transition of QCD  with two massless flavors  within the staggered fermion formulation,
  and this approach is employed in Ref.~\cite{Philipsen:2016hkv} within the Wilson fermion formulation. The approach takes advantage of
 the fact that when the imaginary chemical potential is switched on, the second-order line which separates the first-order region from the crossover region is governed by the
 tricritical scaling law, and the critical exponents  around $am=0$  are known~\cite{Bonati:2012pe,Bonati:2014kpa,Philipsen:2016hkv}.

So far, the investigation to address the nature of the phase transition of QCD  with two massless flavors using this method are implemented through standard gauge and fermion
actions~\cite{Bonati:2014kpa,Philipsen:2016hkv}. Standard KS fermions suffer from taste symmetry breaking at nonzero lattice spacing $a$ \cite{Bazavov:2011nk}.
This taste symmetry breaking can be illustrated by the smallest pion mass taste splitting which is comparable to the pion mass even at lattice spacing $a \sim 0.1fm$ \cite{Bazavov:2009bb}.

In this paper, we aim to investigate the nature of the phase transition of QCD  with two massless flavors  with a one quark-loop Symanzik-improved gauge action
\cite{Symanzik:1983dc,Luscher:1985zq,Lepage:1992xa,Alford:1995hw,Hao:2007iz,Hart:2008sq}
and the HISQ action~\cite{Follana:2006rc}.  The one quark-loop Symanzik-improved gauge action  has a discretization error of $O(\alpha_s^2 a^2, a^4)$,
and the HISQ action completely eliminates the $O(a^2)$ error at tree level by including smeared one-link  and  ``Naik terms''~\cite{Naik:1986bn,Bernard:1997mz}.  Moreover,
the HISQ action  yields the smallest violation of taste symmetry among the currently used staggered actions~\cite{Bazavov:2011nk,Bazavov:2010ru,Cea:2014xva}. These improvements
 are significant on the $N_t=4$ lattice where the lattice spacing is quite large.

The paper is organized as follows.  In Sec.~\ref{SectionLattice},
we define the lattice action with imaginary chemical potential and
the physical observables we calculate.  Our simulation results are
presented in Sec.~\ref{SectionMC},  followed by discussion in
Sec.~\ref{SectionDiscussion}.

\begin{figure*}[t!]
\includegraphics*[width=0.49\textwidth]{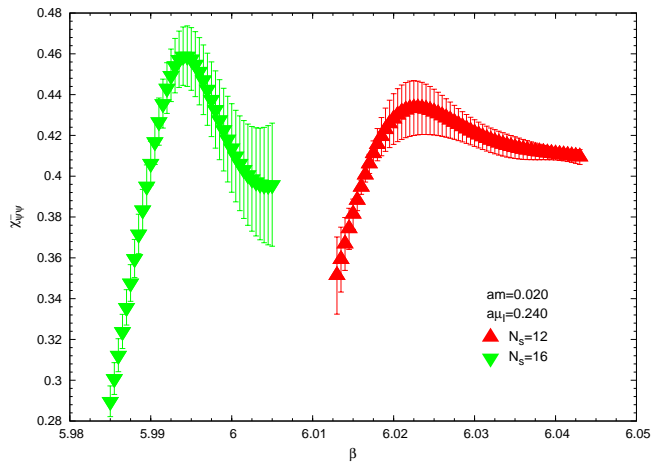}
\caption{\label{fig1} Susceptibility of chiral condensate $\chi_{\bar\psi\psi}$ as a function of coupling $\beta$ at $am=0.020$, $a\mu_I=0.240$ on lattice $12^3\times 4 $  and
$16^3\times 4$.}
\end{figure*}

\begin{figure*}[t!]
\includegraphics*[width=0.49\textwidth]{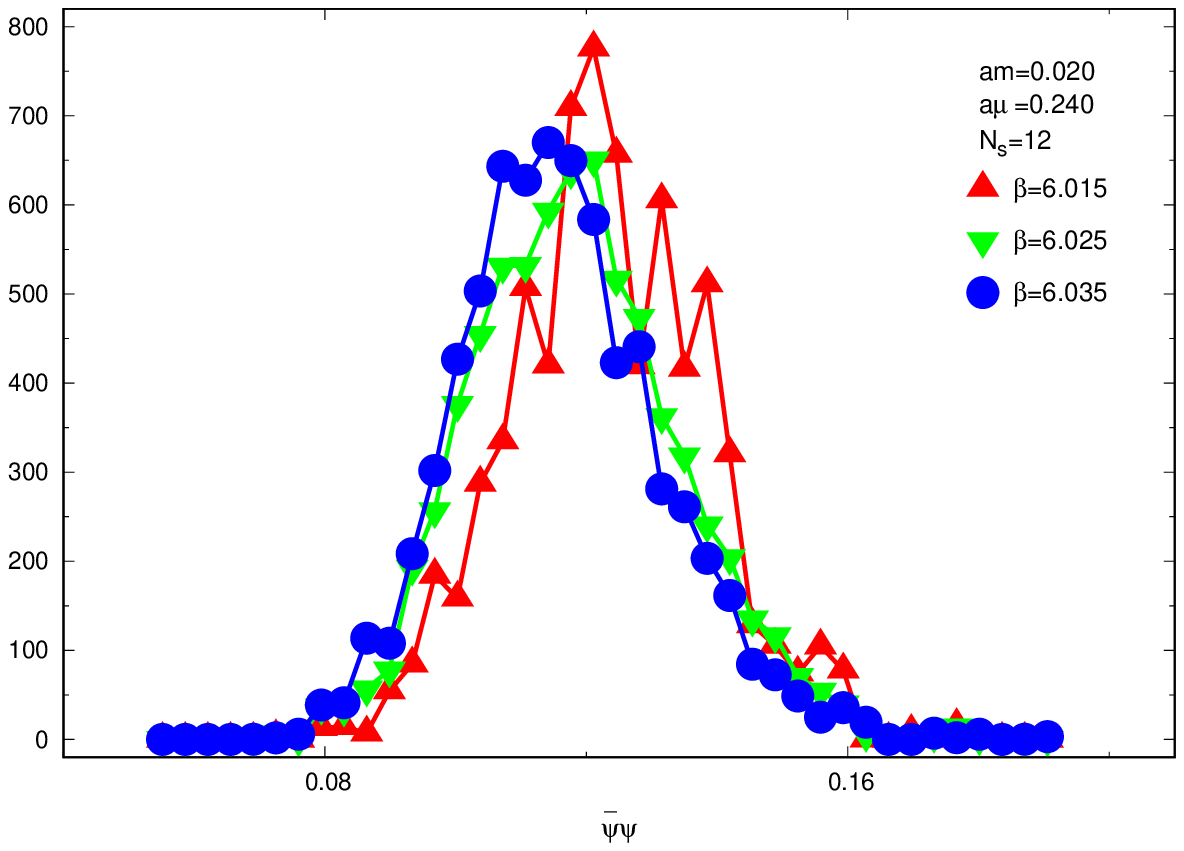}
\caption{\label{fig11} Reweighted distribution of $\bar\psi\psi $ at different $\beta$ at $am=0.020$, $a\mu_I=0.240$ on lattice $N_s=12$. $\beta=6.025 $ corresponds to the
pseudocritical temperature. The horizontal axis represents the value of $\bar\psi\psi$, and the vertical axis stands for the number of
$\bar\psi\psi$,  which is transformed from the probability of corresponding $\bar\psi\psi$. }
\end{figure*}

\begin{figure*}[t!]
\includegraphics*[width=0.49\textwidth]{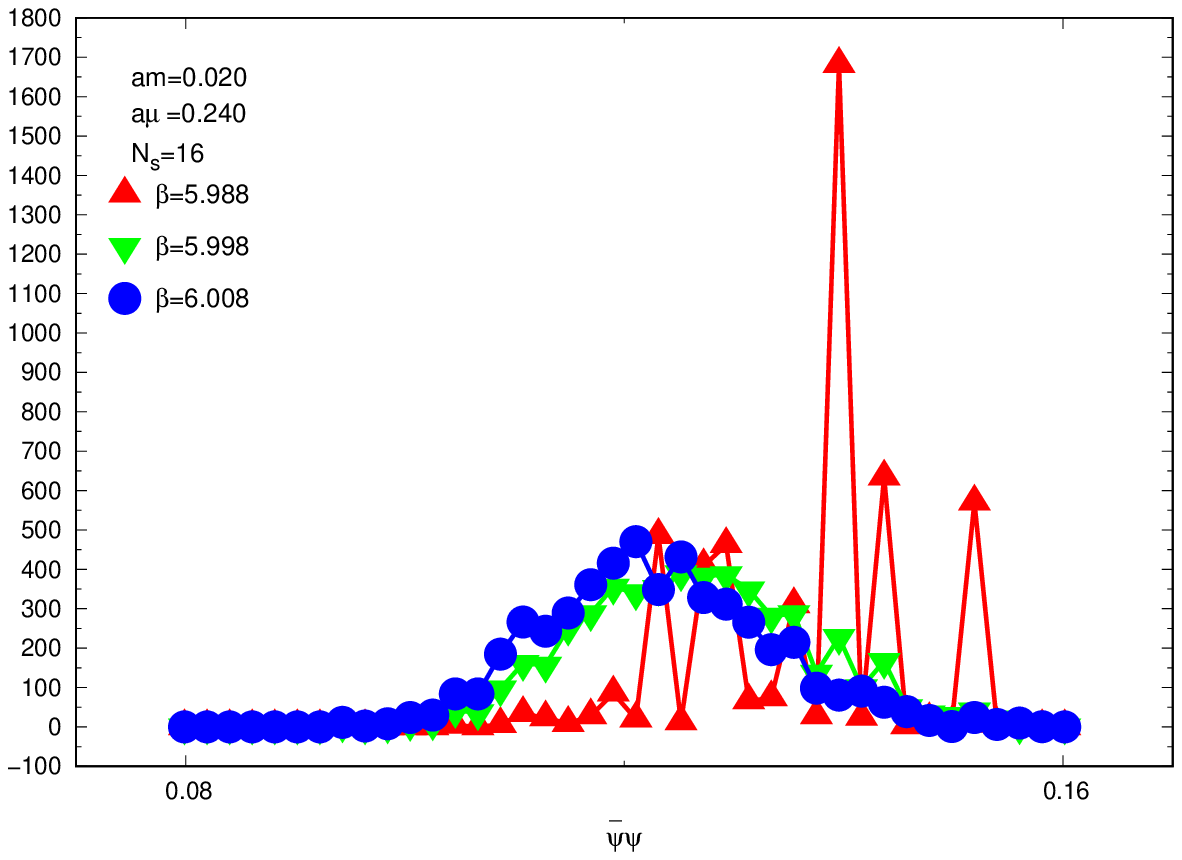}
\caption{\label{fig12} Reweighted distribution of $\bar\psi\psi $ at different $\beta$ at $am=0.020$, $a\mu_I=0.240$ on lattice $N_s=16$ . $\beta=5.998 $ corresponds to the
pseudocritical temperature. The horizontal axis represents the value of $\bar\psi\psi$, and the vertical axis stands for the number of
$\bar\psi\psi$,  which is transformed from the probability of corresponding $\bar\psi\psi$.}
\end{figure*}

\begin{figure*}[t!]
\includegraphics*[width=0.49\textwidth]{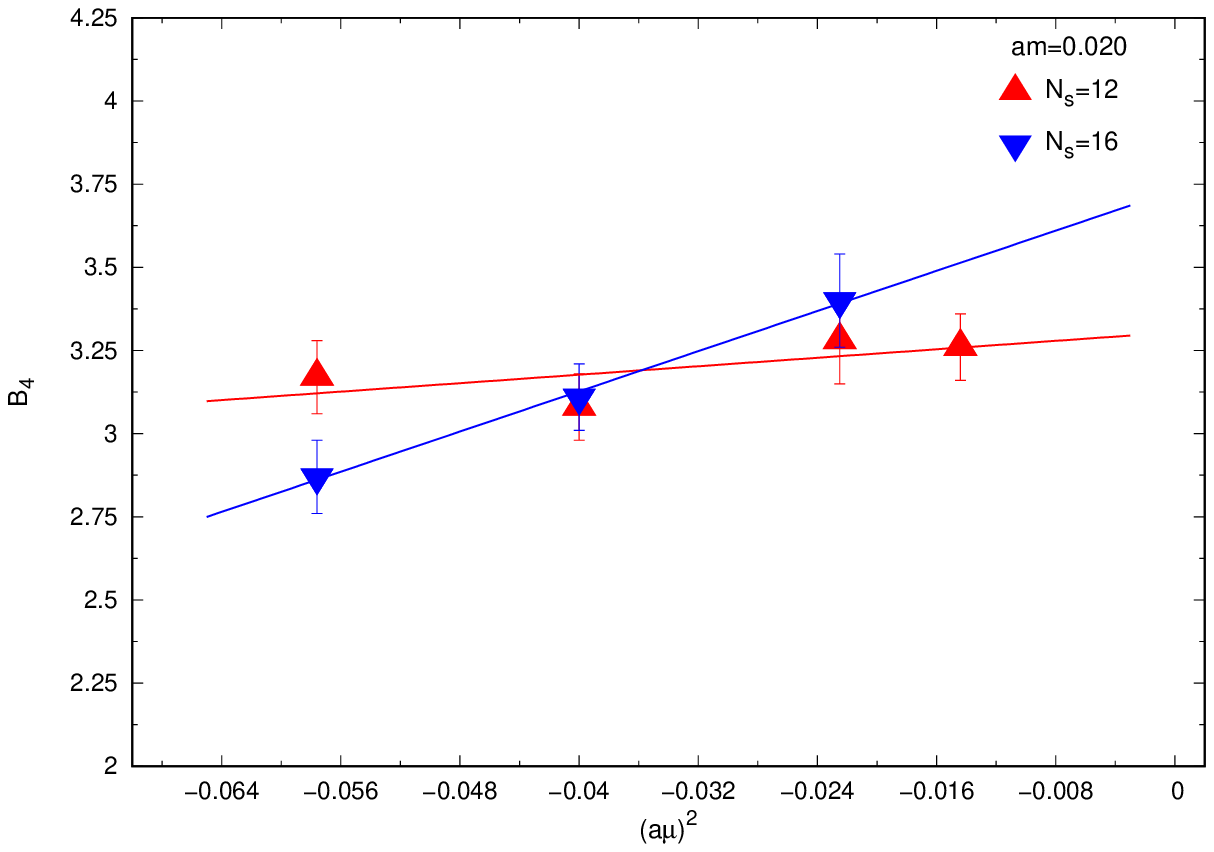}
\includegraphics*[width=0.49\textwidth]{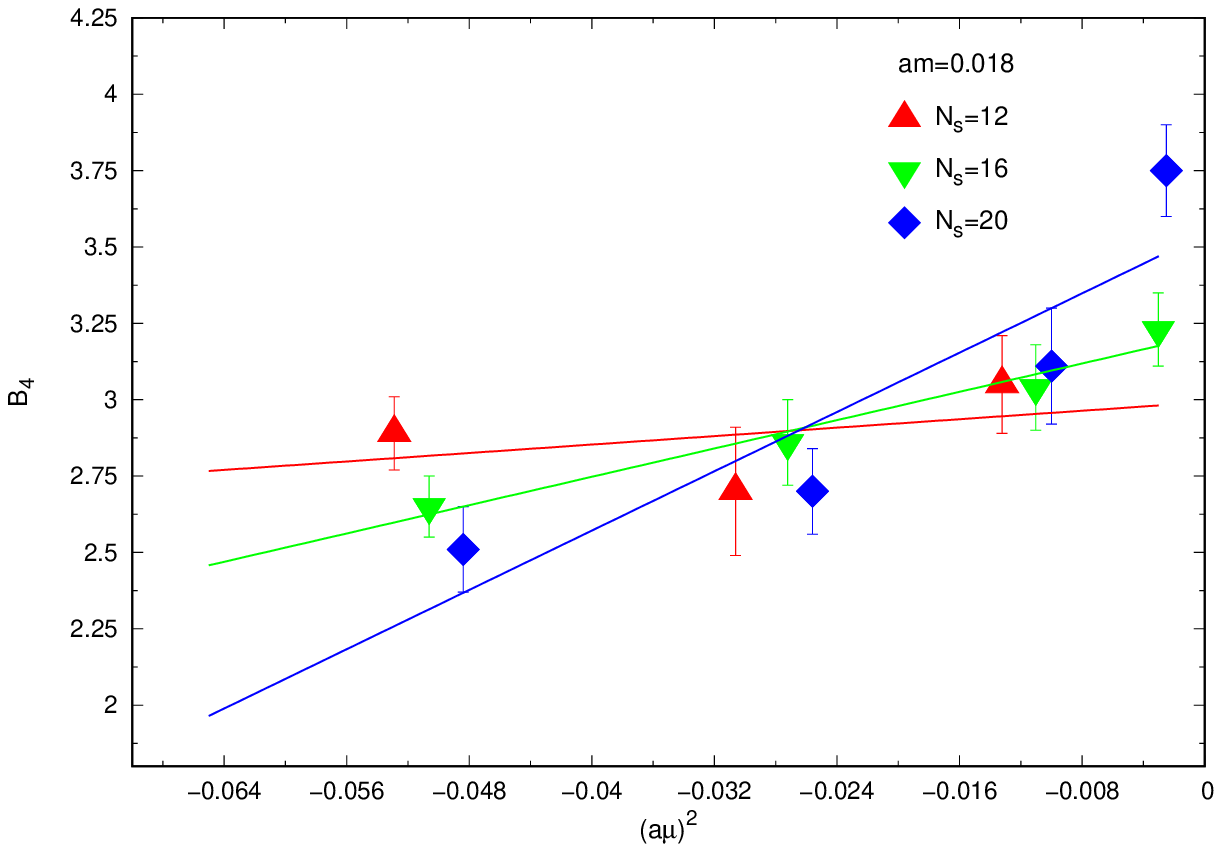}
\caption{\label{fig2}Binder cumulants of $\bar\psi\psi$ at quark masses $am=0.020$ (left panel) and $am=0.018$ (right panel) on different lattice volumes intersect at one point.}
\end{figure*}

\begin{figure*}[t!]
\includegraphics*[width=0.49\textwidth]{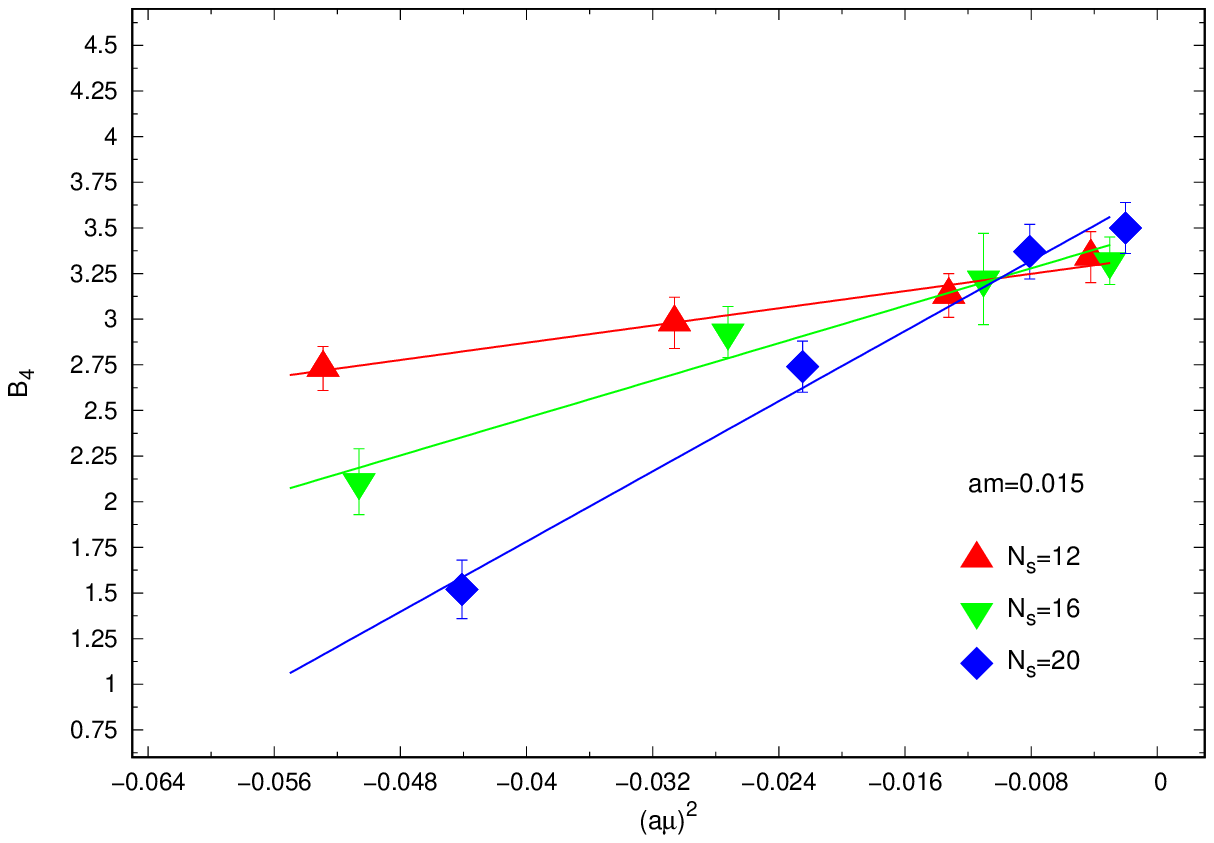}
\includegraphics*[width=0.49\textwidth]{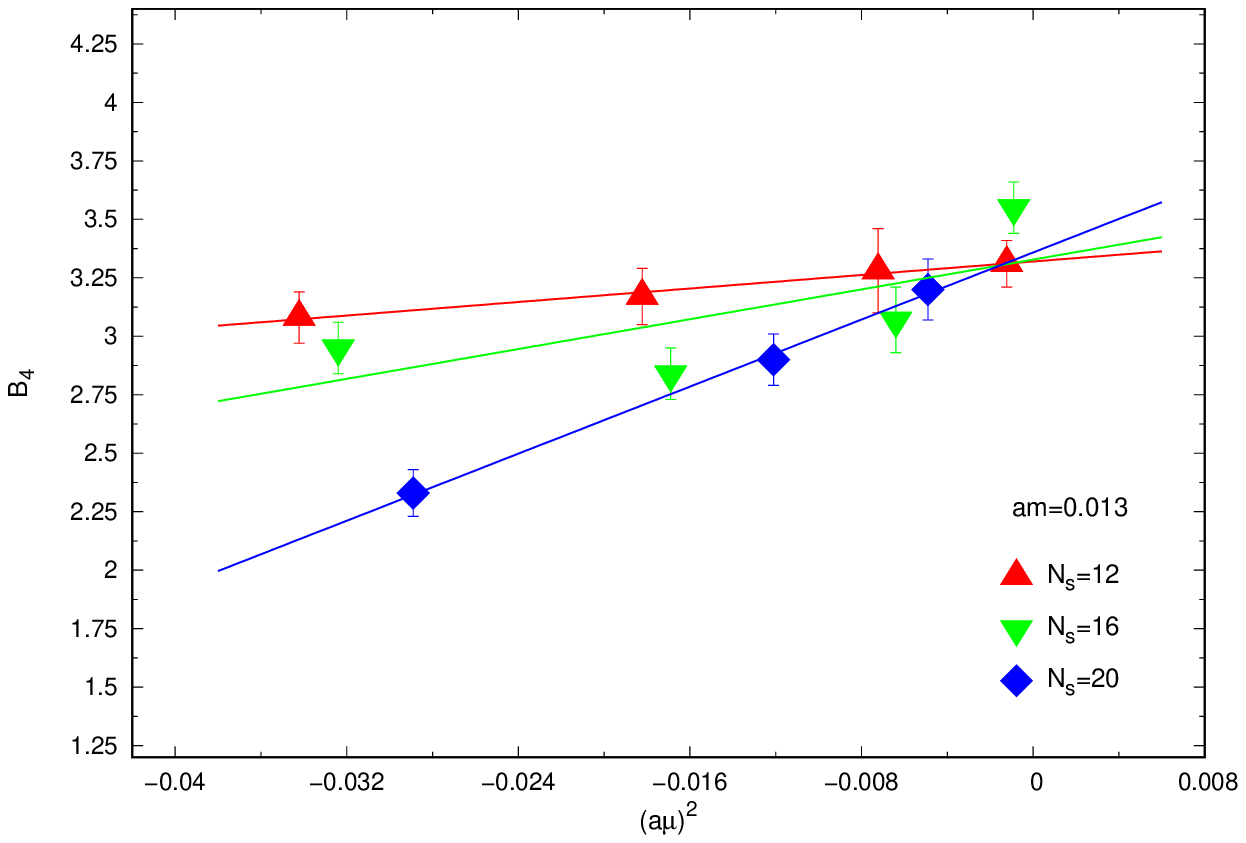}
\caption{\label{fig3} Binder cumulants of $\bar\psi\psi$ at quark masses $am=0.015$ (left panel) and $am=0.013$ (right panel) on different lattice volumes intersect at one point.}
\end{figure*}

\begin{figure*}[t!]
\includegraphics*[width=0.49\textwidth]{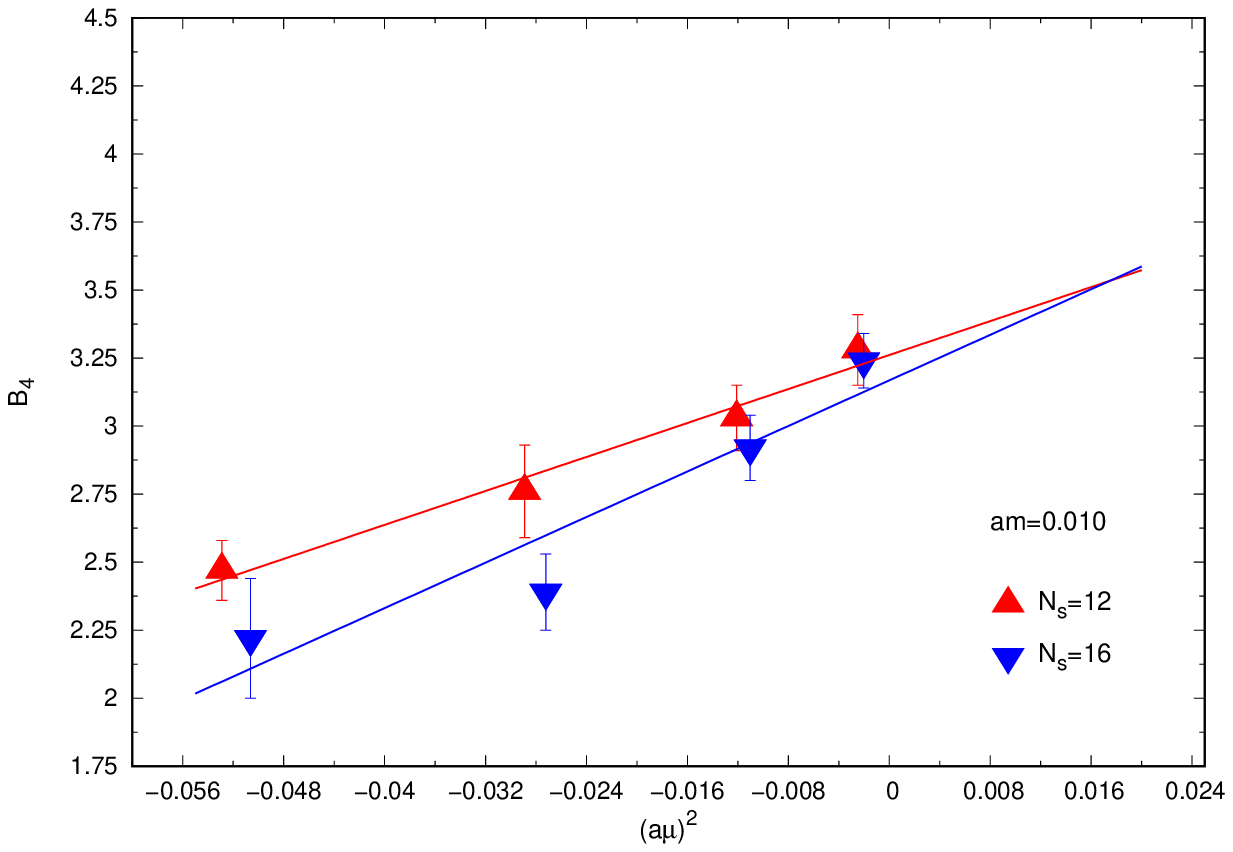}
\caption{\label{fig4} Binder cumulants of $\bar\psi\psi$ at quark mass $am=0.010$  on different lattice volumes intersect at one point.}
\end{figure*}

\section{LATTICE FORMULATION WITH IMAGINARY CHEMICAL POTENTIAL}
\label{SectionLattice} After introducing a pseudofermion field $\Phi$, the partition function of the system can be represented as:
       \begin{eqnarray}
        \label{QCD_partition}
          Z &= &\int [dU][d\Phi^*][d\Phi]e^{-S_g-S_f},
       \end{eqnarray}
where $S_g$ is the Symanzik-improved gauge action, and $S_f$ is the HISQ quark
action with the quark chemical potential $\mu$. Here $\mu=
i\mu_I$ is purely imaginary. For  $S_g$, we use

\newlength{\latlength}
\setlength{\latlength}{1mm}
\setlength{\unitlength}{\latlength}
\ba
  S_g = \beta \left( C_P \sum_{x;\mu<\nu} (1 -  P_{\mu\nu})
      + C_R \sum_{x;\mu < \nu} (1 - R_{\mu\nu}) \right.  \nonumber \\
 \left.      + C_T \sum_{x;\mu<\nu<\sigma} (1 -  T_{\mu\nu\sigma}) \right),
 \label{gaugeaction}
\ea
with  $P_{\mu\nu}, R_{\mu\nu}$,  and $ T_{\mu\nu\sigma}$ standing for $1/3$ of the real part of the trace of $1\times1$, $1\times2$
planar Wilson loops and  $1\times1\times1$ ``parallelogram'' loops, respectivley,
\ba
   P_{\mu\nu} &=& \frac{1}{3} {\rm Re \rm Tr}\
	\raisebox{-4\latlength}{\begin{picture}(10,12)
			\put( 0, 0){\vector( 1, 0){7}}
			\put( 0, 0){\line( 1, 0){10}}
			\put(10, 0){\vector( 0, 1){7}}
			\put(10, 0){\line( 0, 1){10}}
			\put(10,10){\vector(-1, 0){7}}
			\put(10,10){\line(-1, 0){10}}
			\put( 0,10){\vector( 0,-1){7}}
			\put( 0,10){\line( 0,-1){10}}
		\end{picture}}\ ,
\ea
\ba
   R_{\mu\nu} &=& \frac{1}{3} {\rm Re \rm Tr}\
	\raisebox{-4\latlength}{\begin{picture}(20,12)
			\put( 0, 0){\vector( 1, 0){7}}
			\put( 0, 0){\line( 1, 0){10}}
			\put(10, 0){\vector( 1, 0){7}}
			\put(10, 0){\line( 1, 0){10}}
			\put(20, 0){\vector( 0, 1){7}}
			\put(20, 0){\line( 0, 1){10}}
			\put(20,10){\vector(-1, 0){7}}
			\put(20,10){\line(-1, 0){10}}
			\put(10,10){\vector(-1, 0){7}}
			\put(10,10){\line(-1, 0){10}}
			\put( 0,10){\vector( 0,-1){7}}
			\put( 0,10){\line( 0,-1){10}}
			\multiput(10, 1)(0,1){9}{\circle*{0.1}}
		\end{picture}}\ ,
\ea
\ba
   T_{\mu\nu\sigma} &=& \frac{1}{3} {\rm Re \rm Tr}\
	\raisebox{-4\latlength}{\begin{picture}(16,16)
			\put( 0, 0){\vector( 1, 0){7}}
			\put( 0, 0){\line( 1, 0){10}}
			\put(10, 0){\vector( 2, 1){4}}
			\put(10, 0){\line( 2, 1){6}}
			\put(16, 3){\vector( 0, 1){7}}
			\put(16, 3){\line( 0, 1){10}}
			\put(16,13){\vector(-1, 0){7}}
			\put(16,13){\line(-1, 0){10}}
			\put( 6,13){\vector(-2,-1){4}}
			\put( 6,13){\line(-2,-1){6}}
			\put( 0,10){\vector( 0,-1){7}}
			\put( 0,10){\line( 0,-1){10}}
			\multiput(1,0.5)(1,0.5){6}{\circle*{0.1}}
			\multiput(6,4)(0,1){9}{\circle*{0.1}}
			\multiput(7,3)(1,0){9}{\circle*{0.1}}
		\end{picture}}\ ,
\ea

The coefficents $C_P ,C_R$, and $C_T $  are tadpole improved~\cite{Bazavov:2010ru},
\ba
C_P &=& 1.0,  \\
C_R &=& \frac{-1}{20 u_0^2} \left( 1 - \left( 0.6264 - 1.1746 n_f \right) {\rm ln}(u_0) \right),  \\
C_T &=& \frac{1}{u_0^2} \left( 0.0433 - 0.0156 n_f \right) {\rm ln}(u_0).
\ea
with $u_0=(<P_{\mu\nu} >)^{3/4}$.

The HISQ action with pseudofermion field $\Phi$ is
\begin{equation}
S_{f} = \left< \Phi\left|\left[
M^{\dag}[U]M[U]
\right]^{-n_{f}/4} \right|\Phi \right>,
\end{equation}
where the form of $M_{x,y}\left[ U \right]=2m_{x,y}+2\Dslash_{x,y}(U)$  with $2\Dslash_{x,y}(U)$ reading
\begin{eqnarray}
&2\Dslash_{x,y}  = \sum\limits_{\rho=1}^{3}
  \left\{  X_\rho(x) \delta_{x+\hat \rho,y} -  X_\rho^\dagger(x-\hat\rho)\delta_{x-\hat\rho,y} \right\} \nonumber  \\
&
+ \sum\limits_{\rho=1}^{3} \left\{  N_\rho(x) \delta_{x+3\hat \rho,y}  \right.
     - \left. N_\rho^\dagger(x-3\hat\rho)  \delta_{x-3\hat\rho,y} \right\} \nonumber \\
&
 + \left\{   e^{ia\mu_I}X_4(x)\delta_{x+\hat 4,y} -   e^{-ia\mu_I}X_4^\dagger(x-\hat 4)\delta_{x-\hat 4 ,y} \right\}+  \nonumber \\
&
 \left\{ e^{i3a\mu_I} N_4(x)\delta_{x+3\hat 4,y} - 
     e^{-i3a\mu_I}N_4^\dagger(x-3\hat 4)\delta_{x-3\hat 4 ,y} \right\}.
\end{eqnarray}

The Dirac operator $\Dslash$ is constructed from smeared links~\cite{Bazavov:2010ru}.
The fundamental gauge
links are $U_\mu(x)$, the fat links after a level-1  fat7 smearing are $V_\mu(x)$,  the
reunitarized links are $W_\mu(x)$, and the fat links after level-2 asqtad smearing are $X_\mu(x)$.
For simplicity, we use $N_{x,\rho}=W_\rho(x) W_\rho(x+\hat \rho) W_\rho(x+2\hat\rho) $.  The staggered fermion phases are absorbed into the link variables.
$\hat\rho$ and  $\hat 4$ are the unit vectors along $\rho-$direction and $4 $ direction, respectively.

In the study to address the chiral transition, it is natural to choose the chiral condensate as our
observable. The chiral condensate is defined as:
\begin{eqnarray}
X=\bar\psi\psi = \frac{1}{N_s^3N_t}{\rm Tr}(M^{-1}),
\end{eqnarray}
$N_s$ and $ N_t$ are the spatial and temporal extent of lattice, respectively. To simplify notation, we use $X$ to represent
the chiral condensate.
The susceptibility of chiral condensate  is defined as
\begin{eqnarray} \label{chipsi}
\chi_{\bar\psi\psi}= \left\langle( X- \langle  X\rangle)^2\right\rangle.
\end{eqnarray}

We also calculate the Binder cumulant of chiral condensate which is defined as:
\begin{eqnarray}\label{b4psi}
B_4=\left\langle ( X - \langle  X\rangle)^4\right\rangle /
    \left\langle ( X - \langle  X\rangle)^2\right\rangle^2.
\end{eqnarray}

The Binder cumulant of the chiral condensate can be expanded around $a\mu_c$  as ~\cite{Bonati:2014kpa,Philipsen:2016hkv}
 \begin{equation}
 \label{binder_scaling}
 B_4(am, a\mu)  = B_4(am,a\mu_c) + b_1((a\mu)^2  -(a\mu_c)^2)N_s^{1/\nu} + \cdots.
\end{equation}

\section{MC SIMULATION RESULTS}
\label{SectionMC}

Before presenting the simulation results, we describe the computation details.    Simulations are carried out at quark masses $am=0.020,\,$ $0.018,\,$  $0.015,\,$  $0.013,\,$  $0.010$.
The Rational Monte Carlo algorithm \cite{Clark:2003na,Clark:2006wp,Clark:2006fx}
is used to generate configurations. We use different molecular dynamics step sizes for the gauge and fermion parts
of the action, with three gauge steps for each fermion step~\cite{SEXTON}.
 The Omelyan integration algorithm \cite{Takaishi:2005tz,omeylan}
is employed for the gauge and fermion action. For the molecular dynamics evolution,  we use a ninth rational function to approximate $[M^+(U)M(U)]^{-n_f/4}$ for the pseudofermion field.
For the heat bath updating and for computing the action at the beginning and end of the molecular dynamics trajectory,  two tenth rational function are  used to approximate $[M^+(U)M(U)]^{n_f/8}$
and $[M^+(U)M(U)]^{-n_f/8}$, respectively. The step is chosen to ensure the acceptance rate is around $72\%\!\!- \!\!82\% $. Five thousand trajectories of configuration are taken as warmup from
a cold start.    
To fill in  observables   at additional $\beta$ values,
we employ the Ferrenberg-Swendsen reweighting
method~\cite{Ferrenberg:1989ui}.

At a certain pair of the value of quark mass $am$ and chemical potential $a\mu_I$,
we scan the $\beta$ values and  calculate the susceptibility $\chi_{\bar\psi \psi}$ of chiral condensate $\bar\psi \psi$.
The location of peak of susceptibility of the chiral condensate
is interpreted as the transition point. For clarity, we only present the results of $\chi_{\bar\psi\psi}$  on lattice $N_s=12$ and $N_s=16$ at $am=0.020$,
$a\mu_I=0.240$ in Fig~\ref{fig1}.
Similar behavior of $\chi_{\bar\psi\psi}$ can be observed at other couples of
$(am, a\mu_I) $  on different lattice volumes.

To monitor the change of $\bar\psi\psi$ near the pseudotransition point, we present the reweighted distribution of $\bar\psi\psi$ at three
temperatures around the transition   on  $N_s=12$ and $N_s=16$ at $am=0.020$
, $a\mu_I=0.240$
in Figs.~\ref{fig11} and \ref{fig12}, respectively.   The horizontal axis represents the value of $\bar\psi\psi$, and the vertical axis stands for the number of
$\bar\psi\psi$ which is transformed from the probability of corresponding $\bar\psi\psi$.
From Figs.~\ref{fig11} and \ref{fig12}, we can find that the reweighted distribution of $\bar\psi\psi$ does not show the signal of
first-order transition. At other quark masses, similar behavior can be observed.

The results of critical couplings $\beta_c $  and the corresponding
$B_4$ values on  different spatial volumes at different quark masses $am$ are summarized in Table.~\ref{critical_beta}. These $\beta_c$'s are determined from the locations of
peak susceptibility $\chi_{\bar\psi\psi}$ of  chiral condensate $ \bar\psi\psi$.
\begin{table*}[htp]\small
\caption{\label{critical_beta}Results of critical couplings $\beta_c $ obtained by  the locations of peak of $\chi_{\bar\psi\psi}$  and the
$B_4$ values on  different spatial volumes at different quark masses $am$. }
\begin{ruledtabular}
\begin{center}
\begin{tabular}{c|ccc|ccc|ccc|ccc}
    & \multicolumn{3}{c|}{$N_s=8$} & \multicolumn{3}{c|}{$N_s=12$}  & \multicolumn{3}{c|}{$N_s=16$} & \multicolumn{3}{c}{$N_s=20$}  \\
    \hline
$am$  & $a\mu_I$  & $\beta_c$  & $B_4$  & $a\mu_I$  & $\beta_c$  & $B_4$ & $a\mu_I$  & $\beta_c$  & $B_4$   & $a\mu_I$  & $\beta_c$  & $B_4$ \\
\hline
0.010 &   0.040 &   5.998(40) &  3.68(12) &  0.050 &   6.058(20) &  3.28(13) &  0.045 &   6.018(20) &  3.25(10)&  -  &   - &  -\\
      &   0.100 &   6.018(40) &  5.33(12) &  0.110 &   6.048(20) &  3.04(12) &  0.105 &   5.998(20) &  2.93(12)&  -  &   - &  -\\
      &   0.160 &   6.008(20) &  4.03(13) &  0.170 &   5.988(60) &  2.77(17) &  0.165 &   5.998(40) &  2.39(14)&  -  &   - &  -\\
      &   0.220 &   5.988(20) &  4.31(10) &  0.230 &   6.048(20) &  2.48(11) &  0.225 &   6.098(20) &  2.22(22)&  -  &   - &  -\\
0.013 &   0.040 &   5.998(40) &  3.39(15) &  0.035 &   6.008(60) &  3.31(10) &  0.030 &   6.008(60) &  3.55(11) & - &   - &  -\\
      &   0.090 &   5.998(20) &  3.42(11) &  0.085 &   5.968(40) &  3.28(18) &  0.080 &   6.008(10) &  3.08(14) &  0.070 &   6.048(10) &  3.21(13)\\
      &   0.140 &   5.988(30) &  3.43(11) &  0.135 &   5.988(28) &  3.17(12) &  0.130 &   5.964(30) &  2.23(11) &  0.110 &   6.048(30) &  2.91(11)\\
      &   0.190 &   6.048(24) &  3.94(12) &  0.185 &   5.984(34) &  3.09(11) &  0.180 &   6.090(30) &  1.84(11) &  0.170 &   6.098(50) &  2.34(10)\\
0.015 &   0.050 &   6.078(30) &  2.35(22) &  0.065 &   6.018(20) &  3.35(14) &  0.055 &   5.988(30) &  3.33(13) &  0.045 &   6.028(20) &  3.51(14)\\
      &   0.100 &   5.968(40) &  3.78(11) &  0.115 &   6.008(70) &  3.14(12) &  0.105 &   5.988(20) &  3.22(25) &  0.090 &   6.028(20) &  3.37(15)\\
      &   0.160 &   6.028(30) &  3.90(11) &  0.175 &   6.068(110) &  2.98(14) &  0.165 &   6.058(30) &  2.93(14) &  0.150 &   6.098(10) &  2.74(14)\\
      &   0.220 &   5.988(10) &  4.10(13) &  0.230 &   6.028(20) &  2.74(12) &  0.225 &   5.968(100) &  2.12(18) &  0.210 &   5.968(10) &  1.52(16)\\
0.018 &   0.060 &   5.978(30) &  2.20(12) &  0.065 &   6.018(20) &  3.05(10) &  0.055 &   6.018(40) &  3.24(12) &  0.050 &   6.058(10) &  3.75(15)\\
      &   0.110 &   5.958(50) &  2.69(23) &  0.115 &   5.978(50) &  3.05(16) &  0.105 &   6.018(10) &  3.04(14) &  0.100 &   5.978(40) &  3.12(19)\\
      &   0.170 &   5.968(90) &  3.22(12) &  0.175 &   5.968(40) &  2.70(21) &  0.165 &   5.998(30) &  2.86(14) &  0.160 &   5.978(10) &  2.70(14)\\
      &   0.230 &   5.988(30) &  3.61(14) &  0.235 &   6.028(20) &  2.89(12) &  0.225 &   6.018(20) &  2.65(10) &  0.220 &   5.988(40) &  2.51(14)\\
0.020 &   0.060 &   5.998(20) &  2.16(11) &  0.120 &   6.005(30) &  3.26(10) &  0.120 &   6.038(80) &  3.10(16)&  -  &   - &  -\\
      &   0.130 &   5.998(90) &  3.19(17) &  0.150 &   6.078(70) &  3.28(13) &  0.150 &   6.018(40) &  3.40(14)&  -  &   - &  -\\
      &   0.190 &   6.048(40) &  3.31(14) &  0.200 &   6.058(30) &  3.08(10) &  0.200 &   6.078(40) &  3.12(10)&  -  &   - &  -\\
      &   0.250 &   5.998(20) &  3.62(10) &  0.240 &   6.025(10) &  3.18(10) &  0.240 &   5.998(50) &  2.87(11)&  -  &   - &  -\\
\end{tabular}
\end{center}
\end{ruledtabular}
\end{table*}


\begin{table}[htp]
\caption{\label{critical_beta_B4}Results of critical chemical potential $a\mu_c$ at \ different quark masses
obtained by fitting $B_4 = 1.604 + b\left( (a\mu)^2-(a\mu_c)^2 \right)$ to data on lattice $N_s=20$.    }
\begin{ruledtabular}
\begin{center}
\begin{tabular}{c|ccc}

$am$      & $ b $        &   $a\mu_c$   & $ r$-square   \\ \hline
0.013     &  35.84(7)   &   0.2218(2)  &  0.998  \\
0.015     &  48.23(12)) &   0.209(1)   &  0.993    \\
0.018     &  24.28(27)  &   0.282(1)   &  0.899  \\
\end{tabular}
\end{center}
\end{ruledtabular}
\end{table}

After the critical couplings $\beta_c $ and the corresponding
$B_4$ values are obtained, we can monitor their behavior on different lattice spatial volumes at a certain   quark mass. The results are presented in
Figs.~\ref{fig2},\ref{fig3},and \ref{fig4}. From Figs.~\ref{fig2},\ref{fig3}, and \ref{fig4}, we can find that with the decreasing absolute value of the chemical potential, the $B_4$ value increases on lattice
$N_s=12,\,16\,$ and $N_s=20$. On the contrary,  on lattice $N_s=8$, the values of $B_4$ fall with the declining absolute value of chemical potential due to large finite size effect.
So we do not include them in Figs.~\ref{fig2},\ref{fig3}, and \ref{fig4}.
 Nevertheless, at a certain quark mass, we can
find that the $B_4$ values on different lattice volume intersect approximately at one point.

However, from Figs.~\ref{fig2},\ref{fig3}, and \ref{fig4}, we can find that the values of $B_4$ on different lattice sizes approximately intersect
at $B_4(am,a\mu_c) \sim 3$.  We think that it is because of large finite lattice effects.
 To gain some understanding about the result, we fit
expression
\begin{eqnarray} \label{strangefit}
B_4 = 1.604 + b\left( (a\mu)^2-(a\mu_c)^2 \right)
\end{eqnarray}
to  the data on
the $N_s=20$ lattice to get the critical $a\mu_c$. The results are presented in Table~\ref{critical_beta_B4}.

From the results in Table~\ref{critical_beta_B4}, we can see that the critical $a\mu_c$'s on lattice $20^3\times4 $ are approximately  in a reasonable region,  which should be
smaller than 0.262 on the $N_t=4$ lattice. If we use Eq.~(\ref{strangefit}) to fit the data on the smaller lattice, it can be found  that the  critical $a\mu_c$'s are
much larger  than 0.262. Moreover, the $r$-square values in Table~\ref{critical_beta_B4} that are close to 1 show that the fit is good. All these facts imply that
the smaller lattices have significant finite volume effects.

From Fig.~8 in Ref.~\cite{Bonati:2014kpa}, we can see that the value of $B_4$ at the intersection point  is 1.604,  which is consistent with the $Z(2)$ universality class value.
This shows that the finite lattice volume
effects in Ref.~\cite{Bonati:2014kpa} are very small.

\section{DISCUSSIONS}\label{SectionDiscussion}

We have made a simulation in an attempt to understand the nature of the phase transition of QCD with two massless flavors with the one quark-loop Symanzik-improved gauge action and
the HISQ fermion action
by using the method proposed in Ref.~\cite{Bonati:2014kpa} at the quark masses $am=0.020,\, 0.018, \, 0.015, \,
0.013,\, 0.010$.

In our simulation, we found that the Binder cumulants of the chiral condensate on different lattice volumes at one quark mass intersect at one point. The value of $B_4$ at the
intersection point $B_4(am,a\mu_c)$ was renormalization-invariant.
At the quark masses we used , the value of $B_4$ at the
intersection point $B_4(am,a\mu_c)$ was around $3$.

At a nonvanishing  quark mass, an additive and multiplicative renormalization of $\bar\psi\psi$ was needed to define the order
parameter $\bar\psi\psi$ when the scaling property was under consideration~\cite{Ejiri:2009ac,Bazavov:2011nk,Cheng:2007jq}.
Equations.~(9) and (12) in Ref.~\cite{Ejiri:2009ac}, Eq.~(36) in Ref.~\cite{Cheng:2007jq},  and
Eq.~(30) in Ref.~\cite{Bazavov:2011nk}  were used to  subtract the finite quark mass influence on $\bar\psi\psi$.
However, if we start from Eqs.~(\ref{chipsi}) and (\ref{b4psi}) and subtract the finite
quark mass influence from the chiral condensate, then put the subtracted chiral condensate into Eqs.~(\ref{chipsi}) and (\ref{b4psi}),
we think that multiplicative or additive renormalization of $\bar\psi\psi$ would have no effect on the
value of $B_4(am,a\mu_c)$.

If we can detect the $Z(2) $ transition line, the value of $B_4$ at the
intersection point should be  $1.604$~\cite{Bonati:2014kpa,Philipsen:2016hkv}.   However, in our simulation, $B_4(am,a\mu_c) \sim 3$ deviated from the $Z(2) $
universality class value.  If we just used  Eq.~(\ref{strangefit}) to fit the data on $N_s=20$ lattice, we found the fit was good and the
$a\mu_c$'s obtained were reasonable.  So, we think that $B_4(am,a\mu_c) \sim 3$ is because of large finite volume effects as described in Sec.~\ref{SectionMC}.

 Similar behavior was observed in Ref.~\cite{Jin:2017jjp}, in which  Wilson-type fermions were employed to determine the critical point
separating the crossover  from the first-order phase transition region for three-flavour QCD.  In that research,
the value of kurtosis of the chiral condensate at the intersection point deviated from the
universality class value on the $N_t=8, \, 10$ lattice due to finite volume correction.
This observation indicates that simulation with HISQ action along this direction on the $N_s > 20 $ lattice is of great importance.

\begin{acknowledgments}
We thank Gert Aarts, Simon Hands, Chris Allton, and Philippe de Forcrand  for valuable help.
 We modified the MILC Collaboration's
public code~\cite{Milc} to simulate the theory at imaginary chemical
potential.  We used the fortran-90-based multi-precision software~\cite{fortran}.   This work is supported by
the National Natural Science Foundation of China (NSFC)  under Grant No.~11347029, the Key Laboratory of Ministry of Education of China under 
Grant No. QLPL2018P01,  and the National Fund for Studying Abroad of China . The work was carried out at
the National Supercomputer Center in Wuxi and  the National Supercomputer Center in Tianjin.
\end{acknowledgments}


\begin{thebibliography}{99}

\bibitem{Fukushima:2010bq}
  K.~Fukushima and T.~Hatsuda,
  Rept.\ Prog.\ Phys.\  {\bf 74} 014001 (2011). 

\bibitem{Fukushima:2011jc}
  K.~Fukushima,
  J.\ Phys.\ G {\bf 39} 013101 (2012).

\bibitem{Aarts:2015tyj}
  G.~Aarts,
  J.\ Phys.\ Conf.\ Ser.\  {\bf 706}, no. 2, 022004 (2016)
  doi:10.1088/1742-6596/706/2/022004
  [arXiv:1512.05145 [hep-lat]].

\bibitem{Pisarski:1983ms}
  R.~D.~Pisarski and F.~Wilczek,
  Phys.\ Rev.\ D {\bf 29}, 338 (1984).
  doi:10.1103/PhysRevD.29.338


\bibitem{Butti:2003nu}
  A.~Butti, A.~Pelissetto and E.~Vicari,
  JHEP {\bf 0308}, 029 (2003)
  doi:10.1088/1126-6708/2003/08/029
  [hep-ph/0307036].

\bibitem{Pelissetto:2013hqa}
  A.~Pelissetto and E.~Vicari,
  Phys.\ Rev.\ D {\bf 88}, no. 10, 105018 (2013)
  doi:10.1103/PhysRevD.88.105018
  [arXiv:1309.5446 [hep-lat]].


\bibitem{Sato:2014axa}
  T.~Sato and N.~Yamada,
  Phys.\ Rev.\ D {\bf 91}, no. 3, 034025 (2015)
  doi:10.1103/PhysRevD.91.034025
  [arXiv:1412.8026 [hep-lat]].

\bibitem{Ejiri:2009ac}
  S.~Ejiri {\it et al.},
  Phys.\ Rev.\ D {\bf 80}, 094505 (2009)
  doi:10.1103/PhysRevD.80.094505
  [arXiv:0909.5122 [hep-lat]].


\bibitem{Karsch:1994hm}
  F.~Karsch and E.~Laermann,
  Phys.\ Rev.\ D {\bf 50}, 6954 (1994)
  doi:10.1103/PhysRevD.50.6954
  [hep-lat/9406008].




\bibitem{AliKhan:2000wou}
  A.~Ali Khan {\it et al.} [CP-PACS Collaboration],
  Phys.\ Rev.\ D {\bf 63}, 034502 (2001)
  doi:10.1103/PhysRevD.63.034502
  [hep-lat/0008011].

\bibitem{Ejiri:2015vip}
  S.~Ejiri, R.~Iwami and N.~Yamada,
  Phys.\ Rev.\ D {\bf 93}, no. 5, 054506 (2016)
  doi:10.1103/PhysRevD.93.054506
  [arXiv:1511.06126 [hep-lat]].


\bibitem{Burger:2011zc}
  F.~Burger {\it et al.} [tmfT Collaboration],
  Phys.\ Rev.\ D {\bf 87}, no. 7, 074508 (2013)
  doi:10.1103/PhysRevD.87.074508
  [arXiv:1102.4530 [hep-lat]].

\bibitem{Aoki:1998wg}
  S.~Aoki {\it et al.} [JLQCD Collaboration],
  Phys.\ Rev.\ D {\bf 57}, 3910 (1998)
  doi:10.1103/PhysRevD.57.3910
  [hep-lat/9710048].



\bibitem{Bernard:1996iz}
  C.~W.~Bernard {\it et al.},
  Phys.\ Rev.\ Lett.\  {\bf 78}, 598 (1997)
  doi:10.1103/PhysRevLett.78.598
  [hep-lat/9611031].

\bibitem{Iwasaki:1996ya}
  Y.~Iwasaki, K.~Kanaya, S.~Kaya and T.~Yoshie,
  Phys.\ Rev.\ Lett.\  {\bf 78}, 179 (1997)
  doi:10.1103/PhysRevLett.78.179
  [hep-lat/9609022].

\bibitem{Bonati:2014kpa}
  C.~Bonati, P.~de Forcrand, M.~D'Elia, O.~Philipsen and F.~Sanfilippo,
  Phys.\ Rev.\ D {\bf 90}, no. 7, 074030 (2014)
  doi:10.1103/PhysRevD.90.074030
  [arXiv:1408.5086 [hep-lat]].

\bibitem{Philipsen:2016hkv}
  O.~Philipsen and C.~Pinke,
  Phys.\ Rev.\ D {\bf 93}, no. 11, 114507 (2016)
  doi:10.1103/PhysRevD.93.114507
  [arXiv:1602.06129 [hep-lat]].

\bibitem{Cuteri:2017gci}
  F.~Cuteri, O.~Philipsen and A.~Sciarra,
  arXiv:1711.05658 [hep-lat].






\bibitem{DElia:2005nmv}
  M.~D'Elia, A.~Di Giacomo and C.~Pica,
  Phys.\ Rev.\ D {\bf 72}, 114510 (2005)
  doi:10.1103/PhysRevD.72.114510
  [hep-lat/0503030].


\bibitem{Cossu:2013uua}
  G.~Cossu, S.~Aoki, H.~Fukaya, S.~Hashimoto, T.~Kaneko, H.~Matsufuru and J.~I.~Noaki,
  Phys.\ Rev.\ D {\bf 87}, no. 11, 114514 (2013)
  Erratum: [Phys.\ Rev.\ D {\bf 88}, no. 1, 019901 (2013)]
  doi:10.1103/PhysRevD.88.019901, 10.1103/PhysRevD.87.114514
  [arXiv:1304.6145 [hep-lat]].



\bibitem{Aoki:2012yj}
  S.~Aoki, H.~Fukaya and Y.~Taniguchi,
  Phys.\ Rev.\ D {\bf 86}, 114512 (2012)
  doi:10.1103/PhysRevD.86.114512
  [arXiv:1209.2061 [hep-lat]].

\bibitem{Fukugita:1990dv}
  M.~Fukugita, H.~Mino, M.~Okawa and A.~Ukawa,
  Phys.\ Rev.\ D {\bf 42}, 2936 (1990).
  doi:10.1103/PhysRevD.42.2936



\bibitem{Bernard:1999fv}
  C.~W.~Bernard, C.~E.~Detar, S.~A.~Gottlieb, U.~M.~Heller, J.~Hetrick, K.~Rummukainen, R.~L.~Sugar and D.~Toussaint,
  Phys.\ Rev.\ D {\bf 61}, 054503 (2000)
  doi:10.1103/PhysRevD.61.054503
  [hep-lat/9908008].


\bibitem{Meyer:2015wax}
  H.~B.~Meyer,
  PoS LATTICE {\bf 2015}, 014 (2016)
  [arXiv:1512.06634 [hep-lat]].


\bibitem{Ohno:2012br}
  H.~Ohno, U.~M.~Heller, F.~Karsch and S.~Mukherjee,
  PoS LATTICE {\bf 2012}, 095 (2012)
  [arXiv:1211.2591 [hep-lat]].


\bibitem{Dick:2015twa}
  V.~Dick, F.~Karsch, E.~Laermann, S.~Mukherjee and S.~Sharma,
  Phys.\ Rev.\ D {\bf 91}, no. 9, 094504 (2015)
  doi:10.1103/PhysRevD.91.094504
  [arXiv:1502.06190 [hep-lat]].


\bibitem{Brandt:2016daq}
  B.~B.~Brandt, A.~Francis, H.~B.~Meyer, O.~Philipsen, D.~Robaina and H.~Wittig,
  JHEP {\bf 1612}, 158 (2016)
  doi:10.1007/JHEP12(2016)158
  [arXiv:1608.06882 [hep-lat]].

\bibitem{Tomiya:2016jwr}
  A.~Tomiya, G.~Cossu, S.~Aoki, H.~Fukaya, S.~Hashimoto, T.~Kaneko and J.~Noaki,
  Phys.\ Rev.\ D {\bf 96}, no. 3, 034509 (2017)
  Addendum: [Phys.\ Rev.\ D {\bf 96}, no. 7, 079902 (2017)]
  doi:10.1103/PhysRevD.96.034509, 10.1103/PhysRevD.96.079902
  [arXiv:1612.01908 [hep-lat]].


\bibitem{Bazavov:2012qja}
  A.~Bazavov {\it et al.} [HotQCD Collaboration],
  Phys.\ Rev.\ D {\bf 86}, 094503 (2012)
  doi:10.1103/PhysRevD.86.094503
  [arXiv:1205.3535 [hep-lat]].











\bibitem{Bonati:2012pe}
  C.~Bonati, P.~de Forcrand, M.~D'Elia, O.~Philipsen and F.~Sanfilippo,
  PoS LATTICE {\bf 2011}, 189 (2011)
  [arXiv:1201.2769 [hep-lat]].

\bibitem{Bazavov:2011nk}
  A.~Bazavov {\it et al.},
  Phys.\ Rev.\ D {\bf 85}, 054503 (2012)  doi:10.1103/PhysRevD.85.054503  [arXiv:1111.1710 [hep-lat]].  

\bibitem{Bazavov:2009bb}
  A.~Bazavov {\it et al.} [MILC Collaboration],
  Rev.\ Mod.\ Phys.\  {\bf 82} 1349 (2010).


\bibitem{Hao:2007iz}
  Z.~Hao, G.~M.~von Hippel, R.~R.~Horgan, Q.~J.~Mason and H.~D.~Trottier,
  Phys.\ Rev.\ D {\bf 76}, 034507 (2007)
  doi:10.1103/PhysRevD.76.034507
  [arXiv:0705.4660 [hep-lat]].

\bibitem{Symanzik:1983dc}
  K.~Symanzik,
  Nucl.\ Phys.\ B {\bf 226}, 187 (1983).  doi:10.1016/0550-3213(83)90468-6  

\bibitem{Luscher:1985zq}
  M.~Luscher and P.~Weisz,
  Phys.\ Lett.\  {\bf 158B}, 250 (1985).  doi:10.1016/0370-2693(85)90966-9  

\bibitem{Lepage:1992xa}
  G.~P.~Lepage and P.~B.~Mackenzie,
  Phys.\ Rev.\ D {\bf 48}, 2250 (1993)  doi:10.1103/PhysRevD.48.2250  [hep-lat/9209022].  

\bibitem{Alford:1995hw}
  M.~G.~Alford, W.~Dimm, G.~P.~Lepage, G.~Hockney and P.~B.~Mackenzie,
  Phys.\ Lett.\ B {\bf 361}, 87 (1995)  doi:10.1016/0370-2693(95)01131-9  [hep-lat/9507010].  


\bibitem{Hart:2008sq}
  A.~Hart {\it et al.} [HPQCD Collaboration],
  Phys.\ Rev.\ D {\bf 79}, 074008 (2009)
  doi:10.1103/PhysRevD.79.074008
  [arXiv:0812.0503 [hep-lat]].

\bibitem{Follana:2006rc}
  E.~Follana {\it et al.} [HPQCD and UKQCD Collaborations],
  Phys.\ Rev.\ D {\bf 75}, 054502 (2007)
  doi:10.1103/PhysRevD.75.054502
  [hep-lat/0610092].

\bibitem{Naik:1986bn}
  S.~Naik, 1989
  Nucl.\ Phys.\ B {\bf 316} 238 (1989).
\bibitem{Bernard:1997mz}
  C.~W.~Bernard {\it et al.} [MILC Collaboration],
  Phys.\ Rev.\ D {\bf 58} 014503 (1998).





\bibitem{Cea:2014xva}
  P.~Cea, L.~Cosmai and A.~Papa,
  Phys.\ Rev.\ D {\bf 89}, no. 7, 074512 (2014)
  doi:10.1103/PhysRevD.89.074512
  [arXiv:1403.0821 [hep-lat]].




\bibitem{Bazavov:2010ru}
  A.~Bazavov {\it et al.} [MILC Collaboration],
  Phys.\ Rev.\ D {\bf 82}, 074501 (2010)
  doi:10.1103/PhysRevD.82.074501
  [arXiv:1004.0342 [hep-lat]].

\bibitem{Clark:2003na}
  M.~A.~Clark and A.~D.~Kennedy,
  Nucl.\ Phys.\ Proc.\ Suppl.\  {\bf 129} 850 (2004).

\bibitem{Clark:2006wp}
  M.~A.~Clark and A.~D.~Kennedy,
  Phys.\ Rev.\ D {\bf 75}  011502 (2007).
\bibitem{Clark:2006fx}
  M.~A.~Clark and A.~D.~Kennedy,
  Phys.\ Rev.\ Lett.\  {\bf 98}  051601 (2007).


\bibitem{SEXTON}
J.C. Sexton and D.H. Weingarten,
Nucl. Phys. {\bf B380}, 665 (1992).



\bibitem{Takaishi:2005tz}
  T.~Takaishi and P.~De Forcrand,
  Phys.\ Rev.\ E {\bf 73}  036706 (2006).


\bibitem{omeylan}
  I.~P.~Omeylan, I.~M.~Mryglod and R.~Folk,
  Comp.\ Phys.\ Comm.  {\bf 151} 272 (2003).

\bibitem{Ferrenberg:1989ui}
  A.~M.~Ferrenberg and R.~H.~Swendsen,
  Phys.\ Rev.\ Lett.\  {\bf 63}, 1195 (1989).

\bibitem{Cheng:2007jq}
  M.~Cheng {\it et al.},
  Phys.\ Rev.\ D {\bf 77}, 014511 (2008)
  doi:10.1103/PhysRevD.77.014511
  [arXiv:0710.0354 [hep-lat]].

\bibitem{Jin:2017jjp}
  X.~Y.~Jin, Y.~Kuramashi, Y.~Nakamura, S.~Takeda and A.~Ukawa,
  Phys.\ Rev.\ D {\bf 96}, no. 3, 034523 (2017)
  doi:10.1103/PhysRevD.96.034523
  [arXiv:1706.01178 [hep-lat]].


\bibitem{Milc} \url{http://physics.utah.edu/~detar/milc/.}
\bibitem{fortran} \url{http://crd-legacy.lbl.gov/~dhbailey/mpdist/.}



\end{thebibliography}
\end{document}